\documentclass[aps,prb,showpacs,twocolumn,superscriptaddress]{revtex4}
\usepackage{tipa}
\pdfoutput=1
\usepackage{amssymb}
\usepackage{txfonts}
\usepackage{graphicx}

\begin{document}
\title{Influence of the velocity barrier on the massive Dirac electron transport in a monolayer MoS$_{2}$ quantum structure}

\author{X.-J. Hao}
\affiliation{Center for Theoretical
Physics, Department of Physics, Capital Normal University, Beijing
100048, China}
\author{R.-Y. Yuan}
\email{yuanry@cnu.edu.cn} \affiliation{Center for Theoretical
Physics, Department of Physics, Capital Normal University, Beijing
100048, China}
\author{J.-J. Jin}
\affiliation{Center for Theoretical Physics, Department of Physics,
Capital Normal University, Beijing 100048, China}
\author{Y. Guo}
\affiliation{Department of Physics and State Key Laboratory of
Low-Dimensional Quantum Physics, Tsinghua University, Beijing
100084, China}\affiliation{Collaborative Innovation Center of
Quantum Matter, Beijing, China}
\date{{\small \today}}
\begin{abstract}
Using the transfer matrix method, spin- and valley-dependent
electron transport properties modulated by the velocity barrier were
studied in the normal/ferromagnetic/normal monolayer MoS$_{2}$
quantum structure. Based on Snell's Law in optics, we define the
velocity barrier as $\xi=v_{2}/v_{1}$  by changing the Fermi
velocity of the intermediate ferromagnetic region to obtain a
deflection condition during the electron transport process in the
structure. The results show that both the magnitude and the
direction of spin- and valley-dependent electron polarization can be
regulated by the velocity barrier. $-100\%$ polarization of spin-
and valley-dependent electron can be achieved for $\xi>1$, while
$100\%$ polarization can be obtained for $\xi<1$. Furthermore, it is
determined that perfect spin and valley transport always occur at a
large incident angle. In addition, the spin- and valley-dependent
electron transport considerably depends on the length $k_{F}L$ and
the gate voltage $U(x)$ of the intermediate ferromagnetic region.
These findings provide an effective method for designing novel spin
and valley electronic devices.
\end{abstract}

\keywords{velocity barrier, monolayer MoS$_{2}$, Spin, valley,
polarization}

\maketitle

\section {Introduction}

In recent years, approaches to efficiently generate and manipulate spin- or valley-polarized
are the key issues in spintronics and valleytronics. To solve
this problem, various spintronic and valleytronic devices
have been designed that are based on two-dimensional materials, and various
methods (e.g. optical\cite{Cao,Tahir,Heinz1}, electrical
\cite{Li,Sun,Huang,Ye},
magnetic\cite{Yokoyama1,Asgari,Brataas,Che,Yu1} and
temperature\cite{Yu} modulation) have been adopted. Among them, the
spin- and valley-dependent electron transport properties of
monolayer semiconductor transition metal dichalcogenides (TMDCs)
MX$_{2}$ (M = Mo, W; X = S, Se and Te) have attracted considerable
attention\cite{Rashba,Schmidt,Tombros,Ma,Kostya}.

Compared to zero band gap graphene, monolayer semiconductor TMDCs
have a direct band gap ($\approx1.8$ eV) in the visible range. Monolayer MX$_{2}$ has reasonable in-plane
carrier mobility rate, high thermal stability and good compatibility
with standard semiconductor
processes\cite{Splendiani,Mak,Scholes,Law}. At the same time,
monolayer semiconductor TMDCs have strong spin-orbit coupling
(SOC)\cite{Ludwig,MacNeil,Aivazian,Srivastava,Qi,Liu}. For example,
this property can produce a strongly valley-dependent spin split ($\lambda=37.5$ eV) at the top of the
valence band along with the reverse asymmetry in monolayer MoS$_{2}$. In addition,
monolayer semiconductor TMDCs have two intrinsic degrees of freedom
for charge and spin\cite{Tikhonenko,Baibich,Datta}. Similar to
graphene, the first hexagonal Brillouin zone of monolayer
MX$_{2}$ has a pair of degenerate but not equivalent $K$ and $K^{'}$
valleys at the edge of the conduction band and the valence band. The
$K$ and $K^{'}$ valleys are interrelated in the momentum space owing to
time-reversal symmetry, which produces good valley degrees of
freedom at the edges of electrons and holes. A series of novel
phenomena appears based on the valleytronics research, and TMDCs are confirmed to be an ideal energy valleytronics
material\cite{Rycerz,Heinz,Castro,Wu,Dong}.

Owing to these characteristics, TMDCs have broad application prospects in visible light
photoluminescence, high-response photodetectors and field effect
transistors. For instance, spin- and valley-switching effects can be achieved in the p-doped MoS$_{2}$ ferromagnetic/superconducting/ferromagnetic
junction\cite{Majidi}. By applying a ferromagnetic field M or an antiferromagnetic field F in the single or double barriers of MoS$_{2}$, the
transport of ballistic electrons produces different oscillating behaviour\cite{Krstaji}. Meanwhile, it has been demonstrated that the spin and
valley transport can be manipulated effectively by the gate voltage in a normal/ferromagnetic/normal (N/F/N) monolayer MoS$_{2}$ junction\cite{Li}.
These results provide an avenue with different parameters for controlling electron transport in MoS$_{2}$-based devices.

Similar to the regulation of light, an electric field or a magnetic
field, the velocity barrier also has an effective regulation effect
on the electron transport properties in a two-dimensional Dirac
material; this effect has been studied extensively in graphene-based
quantum structures\cite{Raoux,Concha,Krstaji1,Wang} and
ferromagnetic silicone\cite{Ke}. Recently, the effect of the
velocity barrier on spin and valley polarization transport in
monolayer WSe$_{2}$ with strong SOC has been investigated, and a new
path has been opened for high-efficiency spin and valley
polarization in monolayer WSe$_{2}$-based electronic
devices\cite{Qiu}. However, how the velocity barrier affects the
spin- and valley-dependent electron transport properties is still
worth exploring.

In this study, using an N/F/N monolayer MoS$_{2}$ quantum structure,
we investigate the influence of the Fermi velocity barrier on spin-
and valley-dependent electron transport. In this case, the velocity
barrier can be generated by several methods such as by stretching or
extruding the studied material\cite{Castro}, using a
superlattice\cite{Park,Gibertini} or changing the interaction
between surrounding media\cite{Bostwick,Jang}. The results show that
the velocity barrier considerably modulates the transmission and
polarization of spin and valley transport, which is more pronounced
at large angles.

\section {Model AND Formula}

In a given structure, the velocity barrier we consider can be
smoothed by corresponding measures; thus, this structure is an ideal
and effective model\cite{Ke}. We assume that the structure width
$L_{y}$ (in the $y$-direction) is much larger than the ferromagnetic
region length $L$ (in the $x$-direction). Thus, the edge effect of
the structure can be neglected, and the fermions have translation
invariance in the $y$-direction. Therefore, the effect under the
assumption is only the velocity of movement along the $x$-direction.
We determine the Fermi velocity of each region as follows:
\begin{equation}{v_{F}(x)=}
\left \{
\begin{array}{l}
v_{1},  $ $ $ $ $   I, $ $ $ $ $ $ $ $ $x<0,\\
v_{2}, $ $ $ $ $ II$ , $ $ $ $ 0<x<L,\\
v_{1},  $ $ $ $ $ III, $ $ $ x>L,\\
\end{array}
\right.
\end{equation}
where the Fermi velocity in regions I and III is
fixed as $v_{1}=5.3\times10^{5}$m/s\cite{Li1,Tahir1,Li2} and the velocity $v_{2}$ in
region II is adjustable.

\begin{figure}[h]
\includegraphics[width=0.45\textwidth]{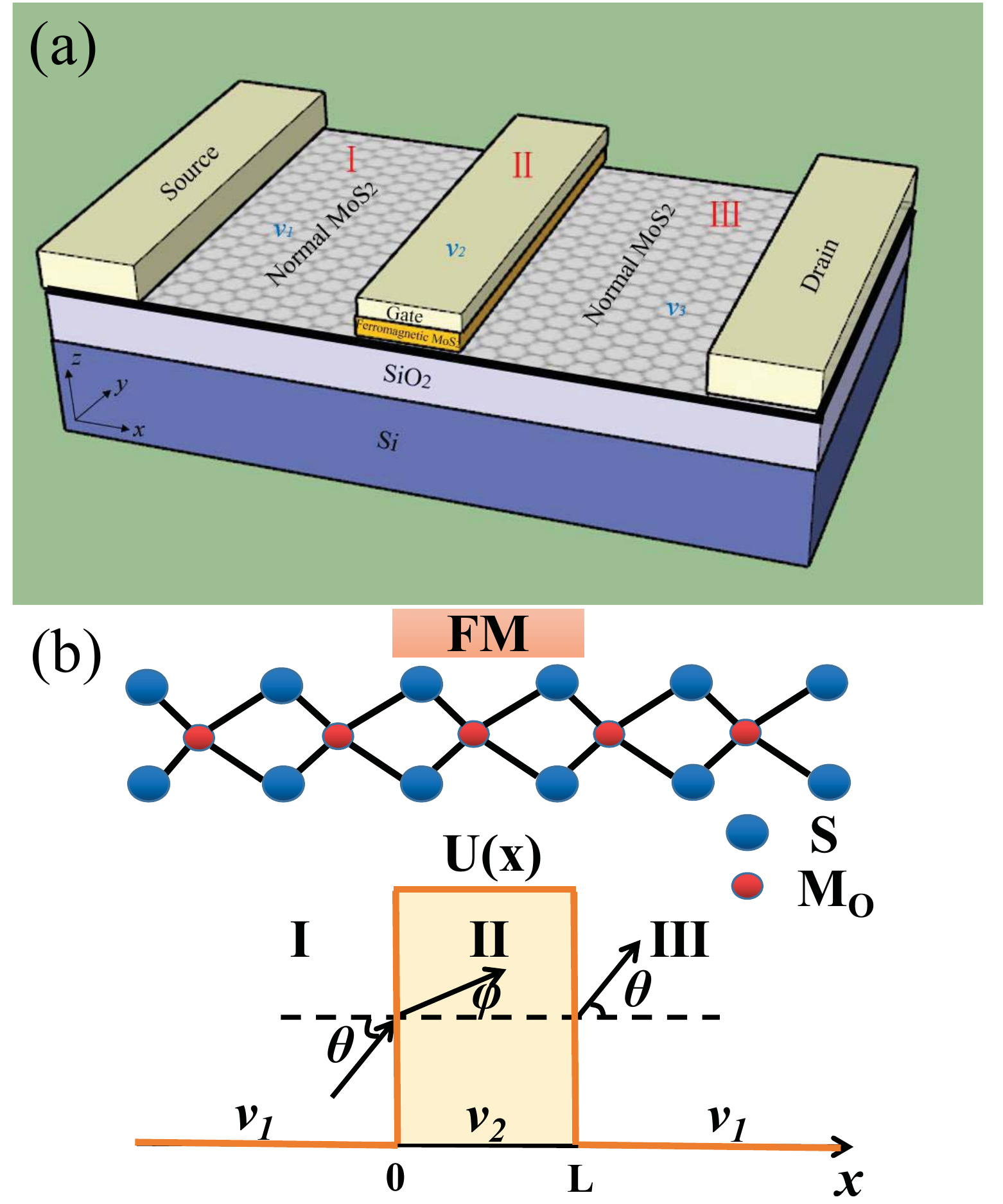}
\caption{\label{Fig1} Schematic diagram of a
normal/ferromagnetic/normal (N/F/N) monolayer MoS$_{2}$ structure
with a velocity barrier. There are a source (left) and a drain
(right) in the structure respectively as shown in Fig. 1(a). A gate
voltage $U(x)$ is added in the ferromagnetic region to generate
electrostatic regulation. The region I and III are normal, $L$ is
the length of the intermediate ferromagnetic region. $\theta$ and
$\varphi$ are the incident angle and the refraction angle
respectively as shown in Fig. 1(b).
 \label{Fig1}}
\end{figure}

According to Fig. 1(b), it is not difficult to determine that the
incident angle and the wave vector of the corresponding region have
the relationship of $\tan\theta=k_{y}/k_{1x}$, and at the same time,
the refraction angle also has the relationship of $\varphi=\tan^{-1}(k_{y}/k_{2x})$.
To extend Snell's Law in optics to quantum mechanics, a
quantum version of Snell's Law is obtained, which provides an
interesting research idea for experimentally adjusting the electron
transport properties by changing the Fermi velocity in different
regions. $\xi=v_{2}/v_{1}$ is defined as the ratio of region II to
region I and III Fermi velocities. The low-energy effective
Hamiltonian of MoS$_{2}$ under tight binding approximation can be
written as
\begin{eqnarray}
H=&\hbar v_{F}(\tau_{z}k_{x}\sigma_{x}+k_{y}\sigma_{y})+\Delta\sigma_{z}\nonumber
\\&+(-\lambda\tau_{z}s_{z}\sigma_{z}+\lambda\tau_{z}s_{z})-s_{z}h+U,
\end{eqnarray}
where $\tau_{z}=\pm1$ represents the $K$ and $K^{'}$ valleys
respectively, $k_{x}$ and $k_{y}$ represent the momentum of the $K(K^{'})$
point, $s_{z}=\pm1$ represents the spin up and down, and
$\sigma_{x}$, $\sigma_{y}$, $\sigma_{z}$ are the Pauli matrix.
$\lambda=37.5$ meV is a spin split induced by SOC at the edge of the
valence band, $\Delta=833$ meV is the band gap caused by the
inversion asymmetry between the orbits $d_{z^{2}}$ and
$(d_{x^{2}-y^{2}} \pm id_{xy})/\sqrt{2}$, and $h$ is the exchange field
added in the ferromagnetic region.

Below, we discuss the wave function for different regions.
First, according to the setup in Fig. 1, we see that in
normal regions, $h=0$ and $U=0$.  Because the wave function follows translational invariance along the $y$-direction, we have
\begin{eqnarray}
\psi(x,y)=\left(
\begin{array}{cc}
\phi_{A}(x,y)  \\
\phi_{B}(x,y) \\
\end{array}\right)=\left(
\begin{array}{cc}
\phi_{A}(x)  \\
\phi_{B}(x) \\
\end{array}\right)e^{ik_{y}\cdot y}.
\end{eqnarray}

Then, by solving the eigenvalue equation $H\psi=E\psi$ with the interface
condition, we obtain that the wave function of region I,
described as
\begin{eqnarray}
\psi_{I}=\frac{1}{E_{N}}\left(
\begin{array}{cc}
\hbar v_{1}k_{1-} \\
E_{M} \\
\end{array}\right)e^{ik_{1x}\cdot x}e^{ik_{y}\cdot y}+ \frac{r}{E_{N}}\left(
\begin{array}{cc}
-\hbar v_{1}k_{1+} \\
E_{M} \\
\end{array}\right)e^{-ik_{1x}\cdot x}e^{ik_{y}\cdot y},
\end{eqnarray}
where $E_{M}=E-\Delta$, $E_{N}^{2}=E_{M}^{2}+(\hbar
v_{1}k_{1})^{2}$, $k_{1+}=\tau_{z}k_{1x}+ik_{y}$,
$k_{1-}=\tau_{z}k_{1x}-ik_{y}$.

Similarly, the wave function of region III is described as
\begin{eqnarray}
\psi_{III}=\frac{t}{E_{N}}\left(
\begin{array}{cc}
\hbar v_{1}k_{1-} \\
E_{M} \\
\end{array}\right)e^{ik_{1x}\cdot x}e^{ik_{y}\cdot y}.
\end{eqnarray}

The wave function of region II is described as
\begin{eqnarray}
\psi_{II}=a^{'}\left(
\begin{array}{cc}
\hbar v_{2}k_{2-} \\
E_{F} \\
\end{array}\right)e^{ik_{2x}\cdot x}e^{ik_{y}\cdot y}+ b^{'}\left(
\begin{array}{cc}
-\hbar v_{2}k_{2+} \\
E_{F} \\
\end{array}\right)e^{-ik_{2x}\cdot x} e^{ik_{y}\cdot y},
\end{eqnarray}
where $E_{F}=E-\Delta-U+s_{z}h$,  $k_{2+}=\tau_{z}k_{2x}+ik_{y}$,
$k_{2-}=\tau_{z}k_{2x}-ik_{y}$. Here, we need to emphasize that $a^{'}, b^{'}, r$ and $t$ are the
spin- and the valley-resolved scattering coefficients.

To more conveniently calculate the transmission of the system, we
use the transfer matrix method to analyze and express the scattering
problem of the given structure. The coordinates on the left and
right boundary surfaces of the barrier are $x=0$ and
$L$ respectively. Considering the conservation of local current at the interface
of the structure
$\overrightarrow{J}(\overrightarrow{r})=v_{F}(\overrightarrow{r})\psi(\overrightarrow{r})\overrightarrow{\sigma}\psi(\overrightarrow{r})$,
the continuous boundary conditions can be imposed as
\begin{eqnarray}
\sqrt{v_{1}}\psi_{I(0^{-})}=\sqrt{v_{2}}\psi_{II(0^{+})},
\sqrt{v_{2}}\psi_{II(L^{-})}=\sqrt{v_{1}}\psi_{III(L^{+})}.
\end{eqnarray}

An expression for the transmission probability can be explicitly
obtained:
\begin{eqnarray}
T_{\tau_{z}, s_{z}}(\theta)=|t|^{2}=|\frac{2k_{2}\cos\varphi E_{M}E_{F}(k_{1-}+\xi k_{1+})e^{-ik_{1}
\cos\theta\cdot L}}{e^{-ik_{2x}\cdot L}P+{e^{ik_{2x}\cdot L}Q}}|^{2},
\end{eqnarray}
where $P=\xi(E_{M}E_{F}k_{1+}k_{2+}+E_{M}^{2}k_{2}^{2})+E_{F}^{2}k_{1}^{2}+E_{M}E_{F}k_{1-}k_{2-}$,
$Q=\xi(E_{M}E_{F}k_{1+}k_{2-}-E_{M}^{2}k_{2}^{2})-E_{F}^{2}k_{1}^{2}+E_{M}E_{F}k_{1-}k_{2+}$. $k_{1}=\sqrt{(E-\tau_{z}s_{z}\lambda)^{2}-(\Delta-\tau_{z}s_{z}\lambda)^{2}}/(\hbar v_{1})$, $k_{2}=\sqrt{(E-\tau_{z}s_{z}\lambda+s_{z}h-U)^{2}-(\Delta-\tau_{z}s_{z}\lambda)^{2}}/(\hbar v_{2})$

The spin- and valley-dependent transmissions are defined as
\begin{eqnarray}
T_{\uparrow(\downarrow)}=\frac{T_{K\uparrow(\downarrow)}+T_{K^{'}\uparrow(\downarrow)}}{2},
\end{eqnarray}
\begin{eqnarray}
T_{K(K^{'})}=\frac{T_{K(K^{'})\uparrow}+T_{K(K^{'})\downarrow}}{2}.
\end{eqnarray}

The spin and the valley polarizations are defined as
\begin{eqnarray}
P_{S}=\frac{T_{\uparrow}-T_{\downarrow}}{T_{c}},
\end{eqnarray}\begin{eqnarray}
P_{V}=\frac{T_{K}-T_{K^{'}}}{T_{c}},
\end{eqnarray}
where
$T_{c}=T_{K\uparrow}+T_{K\downarrow}+T_{K^{'}\uparrow}+T_{K^{'}\downarrow}$.

\section{Results and Discussions}

Fig. 2 shows spin- and valley-dependent transmissions
$T_{\uparrow(\downarrow)}$ and $T_{K(K^{'})}$ as the functions of the
incident angle for different velocity barriers. It can be clearly
seen that the spin- and valley-dependent transmission in the range
of $[-0.5\pi, 0.5\pi]$ is symmetric with the incident angle, namely, $T_{\tau_{z},
s_{z}}(\theta)=T_{\tau_{z}, s_{z}}(-\theta)$. Therefore, we introduce only the change in electron transmission with $\xi$ in the
interval $\theta\in[0, 0.5\pi]$, and it exhibits some interesting
phenomena.

\begin{figure}[h]
\includegraphics[width=0.45\textwidth]{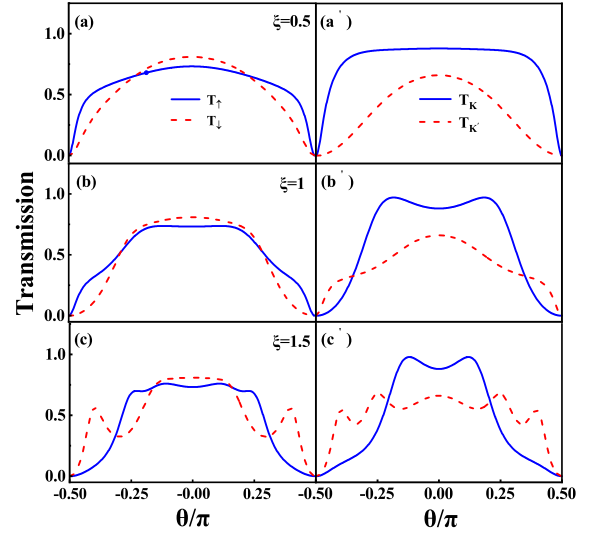}
\caption{\label{Fig2} Spin- and valley-dependent transmission
(a)-(c)$T_{\uparrow(\downarrow)}$ and (a$^{'}$)-(c$^{'}$)
$T_{K(K^{'})}$ as the functions of incident angle $\theta$ with
different velocity barrier ratio $\xi=0.5, 1, 1.5$. The other
parameters are $L=7$nm, $U=-2.0\Delta, E=1.2\Delta, h=0.2\Delta$.
 \label{Fig2}}
\end{figure}

In general, with an increase in the velocity barrier ratio $\xi$
shown in Fig. 2, the oscillations of the curves for spin- and
valley-dependent electron transmission are both strengthened. This
phenomenon indicates that the greater the velocity barrier ratio
$\xi$, the more obvious the resonant effect. This property is
confirmed by the transmission expression in Eq.(8). The
oscillations, which are not periodic, originate from the unequal
factors $e^{-ik_{2x}\cdot L}$ and $e^{ik_{2x}\cdot L}$ of the
transmission coefficient. In addition, the velocity barrier ratio
$\xi$ plays an important role in the transmission expression.
Therefore, with the appropriate adjustment of the Fermi velocity,
the quantum structure has potential applications in resonant
tunneling devices.

For $\xi=0.5$ as shown in Figs. 2(a) and (a$^{'}$), the curves of
spin- and valley-dependent electron transmission change smoothly
with an increase in the incident angle. The spin-up electron
transmission is smaller than the spin-down electron transmission
within small incident angles ($\theta<0.25\pi$), as shown in Fig.
2(a), which indicates that the polarization of spin-down electrons
occurs. When the incident angle increases ($\theta>0.25\pi$), the
spin-down electron transmission decreases faster than the spin-up
electron transmission, which makes the spin-down electron
transmission is smaller than the spin-up electron transmission. It
is demonstrated that the reversal effect of spin electron
polarization can be achieved at $\theta=0.25\pi$. However, for
$\xi=0.5$, the valley-dependent transmission versus the incident
angle is specific to the spin-dependent transmission. The $K$ valley
transmission is always higher than the $K^{'}$ valley transmission
in the entire internal $\theta\in[0, 0.5\pi]$ (see Fig. 2(a$^{'}$)).
Moreover, when the incident angle increases gradually, the $K^{'}$
valley transmission monotonously decreases exponentially. At the
same time, the $K$ valley transmission remains nearly $100\%$ for
$\theta\in[0, 0.35\pi]$ while decreases rapidly when the incident
angle $\theta>0.35\pi$.

When the velocity barrier ratio $\xi=1.0$, as shown in Fig. 2(b),
the curves of the spin electron transmission have slight
oscillations, and the incident angle of polarization inversion is
shifted backward ($\theta=0.30\pi$). Interestingly, for $\xi=1.0$,
the inversion of polarization is also presented in the valley
transmission curves. As shown in Fig. 2(b$^{'}$), with an increase
in the incident angle, the $K$ valley transmission is enhanced
slightly and then suppressed sharply, and the $K^{'}$ valley
transmission decreases monotonously. $T_{K}>T_{K^{'}}$ within the
range of $[0, 0.35\pi]$. When the incident angle $\theta>0.35\pi$,
the $K$ valley transmission decreases faster than the $K^{'}$ valley
transmission, which results in $T_{K}<T_{K^{'}}$, and the
polarization of the electron for the $K^{'}$ valley occurs. In
addition, we also need to note that $100\%$ transmission for the $K$
valley can be still obtained only when the incident angle
$\theta=0.15\pi$.

With an increase in the Fermi velocity barrier ratio to 1.5, as
shown in Figs. 2(c) and (c$^{'}$), the spin- and valley-dependent
electron transmission oscillates clearly. Fig. 2(c) showns that spin
polarization reversals can be achieved respectively at
$\theta=0.12\pi$ and $\theta=0.30\pi$. The polarization of the spin-down
electron can be obtained in the internal of $[0, 0.12\pi]$ and
$[0.30\pi, 0.50\pi]$, while the polarization of the spin-up electron
occurs within the range of $[0.12\pi, 0.30\pi]$. However, the curves
of the valley electron transmission have only one cross point. The
valley polarization inversion is obtained at $\theta=0.2\pi$, and the
angle of the inversion is clearly shifted forward, as shown in Fig.
2(c$^{'}$). Therefore, these results imply that for the large
velocity barrier ratio compared with the valley-dependent electron
transmission, the spin-dependent electron transmission is more
sensitive to the incident angle. Moreover, it is shown that
$T_{K}\neq1$ at normal incidence $\theta=0$, which indicates that
perfect transmission does not exist only at normal incidence. Here,
we adjust the incident angle to $0.12\pi$, and $100\%$ transmission
of the $K$ valley still exists.

\begin{figure}[h]
\includegraphics[width=0.45\textwidth]{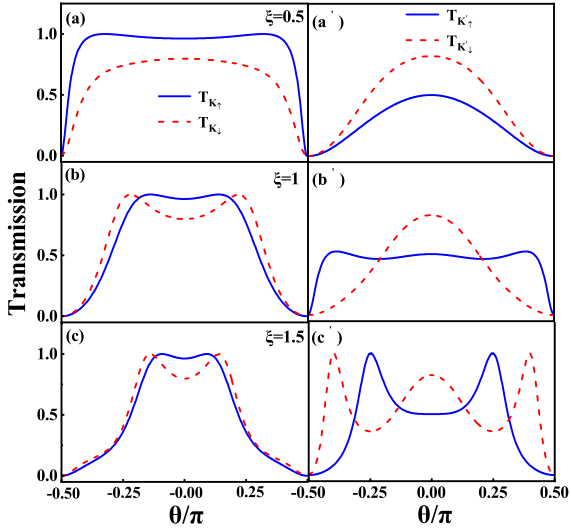}
\caption{\label{Fig3} Spin- and valley-resolved transmission (a)-(c) $T_{K\uparrow(\downarrow)}$ and (a$^{'}$)-(c$^{'}$)
$T_{K^{'}\uparrow(\downarrow)}$ versus incident angle $\theta$ with different
velocity barrier ratio $\xi=0.5, 1, 1.5$. The other parameters are
the same as Fig. 2.
 \label{Fig3}}
\end{figure}

To more clearly illustrate how to produce the reversal effect of
polarization by modulating the incident angle, we plot the curves of
spin- and valley-resolved transmission as the functions of incident
angle $\theta$ with different velocity barriers, as shown in Fig. 3.
According to the transmission expression in Eq.(8), we calculated
the values of the incident angle where the reversal effect of
polarization appears. For instance, the incident angles
$\theta=0.25\pi$ for $\xi=0.5$ in the Fig. 2(a) and for
$\theta=0.30\pi$ for $\xi=1$ in Fig. 2(b) are achieved. It is
demonstrated that the change of the curves in Fig. 3 is generally
the same as the results calculated from the transmission expression.
We determine that when the incident angle $\theta<0.25\pi$, as shown
in Fig. 3(a), the transmission of spin electron for the $K$ valley
is clearly unchanged. While the incident angle $\theta>0.25\pi$, the
transmission of spin-up electron for the $K$ valley increases
slightly to $100\%$, then decreases rapidly to zero. At the same
time, the transmission of spin-down electron for the $K$ valley
decreases to zero. $T_{K\uparrow}>T_{K\downarrow}$ in the entire
incident angle range. However, for spin electron for the $K^{'}$
valley, all curves of transmission decrease to zero with different
amplitudes, as shown in Fig. 3(a$^{'}$), and
$T_{K^{'}\downarrow}>T_{K^{'}\uparrow}$ in the entire range of
incident angles. Taking the spin and valley indices in Eqs.(9) and
(10) into account, it is determined that at the incident angle of
$\theta=0.25\pi$, the reversal effect of polarization occurs, as
shown in Fig. 2(a), for the velocity ratio of $\xi=0.5$. At other
incident angle, the physical origin of the reversal effect of
polarization is similar to the one mentioned above.

\begin{figure}[h]
\includegraphics[width=0.45\textwidth]{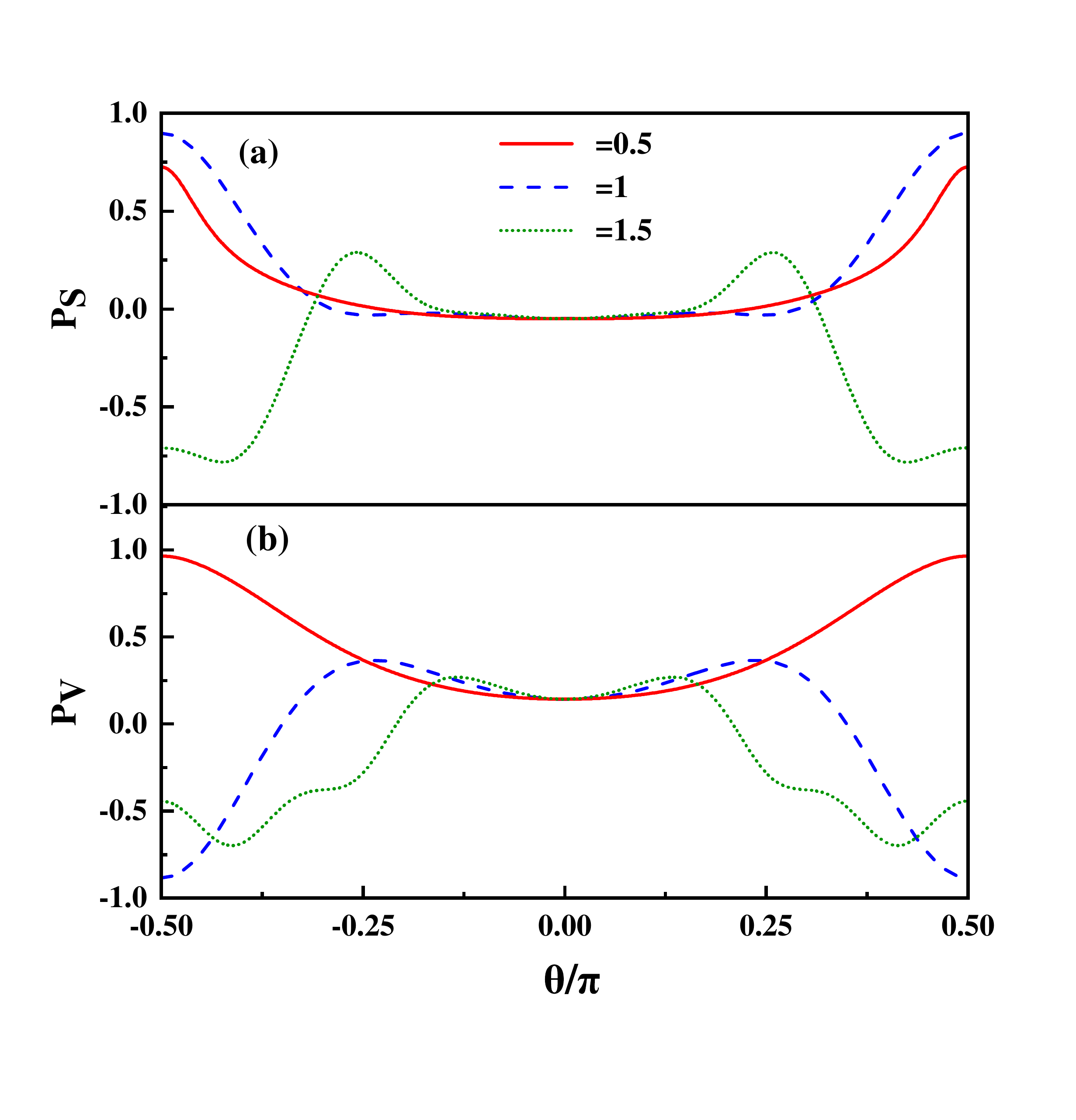}
\caption{\label{Fig4} Spin and valley polarization $P_{S}$(a) and
$P_{V}$(b) as the functions of incident angle $\theta$ with different
velocity barrier ratio $\xi=0.5, 1, 1.5$. The other parameters are
the same as Fig. 2.
 \label{Fig4}}
\end{figure}

To clearly demonstrate the modulation of the velocity barrier ratio
$\xi$ on spin- and valley-dependent electron transport, in Fig. 4,
we provide the spin and valley polarization $P_{S}$(a) and
$P_{V}$(b) as the functions of the incident angle $\theta$ with a
different velocity barrier ratio.

For $\xi=0.5$ and $1.0$ with an increase in the incident angle, the
spin polarization increases slightly in the positive direction. For
$\xi=1.5$, it is clearly observed that spin polarization increases
in the negative direction for $\theta>0.30\pi$. Furthermore, at the
large incident angle, the $100\%$ polarization of spin electron can
be achieved for $\xi=1$, as indicated by the dashed (blue) line,
while $-100\%$ polarization for spin electron can be obtained [the
dotted (green) line] for $\xi=1.5$ in Fig. 4(a). However, compared
to the spin-dependent electron polarization, it is determined that
to achieve the $100\%$ polarization for the valley electron, $\xi=0.5$
is appropriate [the solid (red) line], as shown in Fig. 4(b).
Meanwhile, it is determined that $-100\%$ valley polarization can be
obtained for $\xi=1.0$, as shown by the dashed (blue) line. These
results further demonstrate that the polarization of spin- and
valley-dependent electron can be modulated effectively by the
velocity barrier.

\begin{figure}[h]
\includegraphics[width=0.45\textwidth]{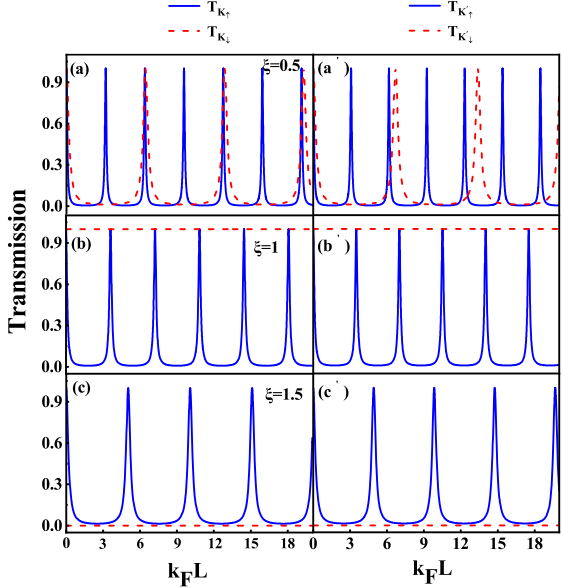}
\caption{\label{Fig5} Spin- and valley-resolved transmission (a)-(c) $T_{K\uparrow(\downarrow)}$ and (a$^{'}$)-(c$^{'}$)
$T_{K^{'}\uparrow(\downarrow)}$ versus length $k_{F}L$ with
different velocity barrier ratio $\xi=0.5, 1, 1.5$, where $k_{F}=E/(\hbar v_{F}), E=1.2\Delta$. The other
parameters are $\theta=0.48\pi, U=-0.2\Delta,
h=0.2\Delta$.
 \label{Fig5}}
\end{figure}

Fig. 4 clearly shows the influence of the velocity barrier on
polarization for the spin- and valley-dependent electron can be
achieved readily for large incident angles. Thus, we investigate the
transport properties of spin- and valley-resolved electron versus
length $k_{F}L$ at the larger incident angle $\theta=0.48\pi$, as
shown in Fig. 5. The left column is the $K$ valley and the right
column is the $K^{'}$ valley. First, line-type resonance occurs for
spin- and valley-dependent electron transmission. Second, the
responses of transmission for spin electron at the $K$ and $K^{'}$
valleys are almost the same for the length $k_{F}L$ for the
corresponding velocity barriers. In addition, it is also shown that
the number of line-type resonant peaks tends to decrease with a
gradual increase in the velocity barrier ratio $\xi$. This result
indicates that fewer spin- and valley-dependent electron will pass
through the barrier.

For the velocity barrier ratio of $\xi=0.5$ shown in Fig. 5(a), the
oscillation frequency versus $k_{F}L$ and amplitude of the line-type
resonance almost overlap, and it is difficult to achieve the spin
electron polarization for the $K$ valley. However, spin electron
flittering at the $K^{'}$ valley could be achieved at some specific
values of $k_{F}L$, as shown in Fig. 5(a$^{'}$). For the velocity
barrier ratio $\xi=1$, as shown in Figs. 5(b) and (b$^{'}$), the
spin-down electron of the $K$ and $K^{'}$ valleys pass perfectly
through a given device without reflection. This indicates that the
spin-down electron transport of the two valleys is independent of
the structure length. Moreover, when $k_{F}L$ is near the values of
3, 7, 11, 15 and 18, the transmission of the spin-up electron in
the two valleys suddenly sharply increases to $100\%$, while at the
other values of $k_{F}L$, $T_{K\uparrow}$ and $T_{K^{'}\uparrow}$
are almost 0. However, when the velocity barrier ratio $\xi$
increases to 1.5, the transmission of spin-down electron in two
valleys is suppressed completely. The $100\%$ transmission of
spin-up electron for two valleys can be acquired around the values
of 5, 10 and 15. Meanwhile, the spin-up electron has a significant
change in the number of line-type resonance peaks with an increase
in the length $k_{F}L$.

Therefore, line-type resonance can turn up
under the action of the velocity barrier, which is essential for the
spin and valley filtering effects at the larger incident angle. This
property can be achieved from the line-type resonance conditions
$cos\theta\sim0$ and $k_{2}cos\varphi\cdot L=n\pi$. Then we can
obtain equation $(n\pi\hbar
v_{F})/L=\sqrt{(E-U-\lambda\tau_{z}s_{z}+s_{z}h)^{2}-(E-\lambda\tau_{z}s_{z})^{2}}$
with the translation invariance in the $y$-direction. The
above-mentioned equation shows that only some massive Dirac
electrons equipped with specific energy can tunnel into the
interfaces and are strengthened by the line-type resonance at the
large incident angle. In addition, based on the transmission
expression in Eq.(8), we determined that when the velocity barrier
ratio increases, the number of line-type resonant peaks decreases.
Otherwise, these interesting phenomena indicate that the spin-down
electron can be filtered out within several particular intervals of
length $k_{F}L$ for $\xi=1$ in the $K$ and $K^{'}$ valleys.
Nevertheless, the filtering effect of the spin-up electron can be
achieved at the specified length $k_{F}L$ for $\xi=1.5$.
Consequently, a sensitive current switching device that is based on
the spin and valley indices can be theoretically designed in the
considered model.

\begin{figure}[h]
\includegraphics[width=0.45\textwidth]{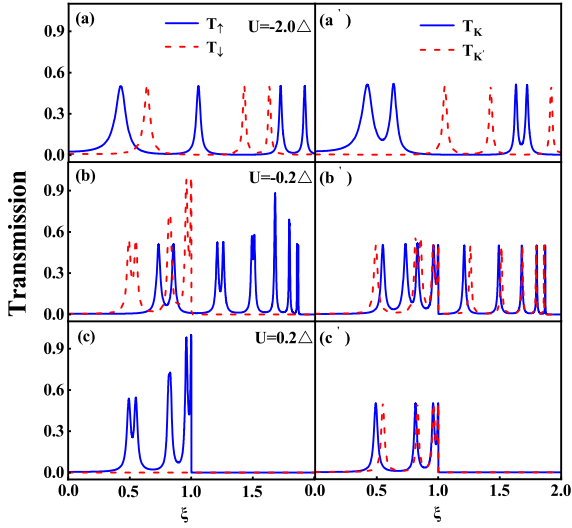}
\caption{\label{Fig6} Spin- and valley-dependent transmission
$T_{\uparrow(\downarrow)}$, $T_{K(K^{'})}$ as the functions of
velocity barrier ratio $\xi$ with different gate voltage $U$ for $\theta=0.48\pi$. Here,
the parameters are $U=-2.0\Delta$ in (a) and (a$^{'}$);
$U=-0.2\Delta$ in (b) and (b$^{'}$); $U=0.2\Delta$ in (c) and
(c$^{'}$). The other parameters are
the same as Fig. 2.
 \label{Fig6}}
\end{figure}

Next, we focus on the the spin- and valley-dependent electron
transmissions as the functions of the velocity barrier with various
gate voltages. It is illustrated that the transmission of spin and
valley electron is not too high with the negative voltage of
$U=-2.0\Delta$, as shown in Figs. 6(a) and (a$^{'}$).
$T_{\uparrow(\downarrow)}$ and $T_{K(K^{'})}$ are always below 0.6.
For $U=-0.2\Delta$, the electron transmission curves of the two
valleys almost overlap. Thus, the effect of the velocity barrier on
the electron transmission of the $K$ and $K^{'}$ valleys is almost
meaningless, as shown in Fig. 6(b$^{'}$). However, the
spin-dependent electron can achieve perfect filtering for $\xi>1$
shown in Fig. 6(b). This occurs because the spin-down electron is
always suppressed for $\xi>1$, and the transmission of the spin-up
electron is approximately $100\%$ at some special velocity ratios
ratios. When the voltage is $U=0.2\Delta$, the number of line-type
resonant peaks is considerably reduced, as shown in Figs. 6(c) and
(c$^{'}$). Of course, it is clear that the positive voltage has a
better spin filtering effect on electron because the transmission of
spin-down electron in this case is independent of the velocity
barrier ratio $\xi$, and $T_{\downarrow}$ is always 0 at the given
configuration. Thus, for $\xi<1$, the spin-up current can be
obtained by adjusting the corresponding velocity barrier ratio
$\xi$; the transmission of spin-up electron is close to $100\%$ at
$\xi=0.85$ and 1.0. These behaviors have some important practical
significance in future spintronic and valleytronic devices.

\begin{figure}[h]
\includegraphics[width=0.45\textwidth]{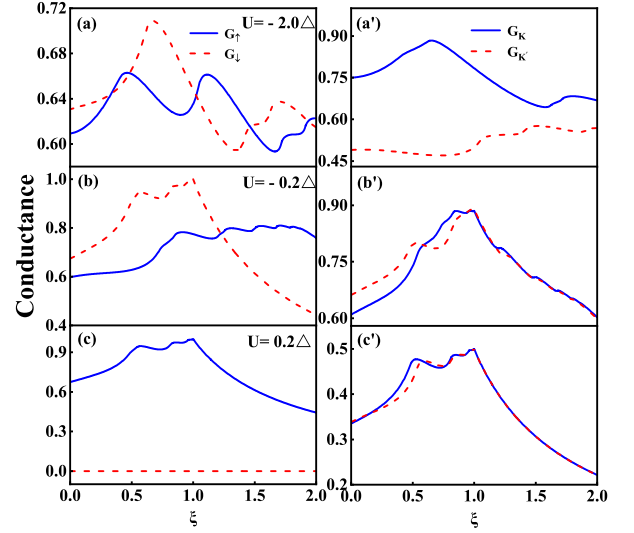}
\caption{\label{Fig7} Spin- and valley-dependent conductance
$G_{\uparrow(\downarrow)}$ and $G_{K(K^{'})}$ as the functions of
velocity barrier ratio $\xi$ with different gate voltage $U$. The other parameters are
the same as Fig. 2.
 \label{Fig7}}
\end{figure}

It is known that conductance is an important and measurable index
for the experimental evaluation of spin-valley
devices, and it is given as
$G_{\tau_{z},s_{z}}=\frac{1}{2}\int^{\pi/2}_{-\pi/2}T_{\tau_{z},s_{z}}cos\theta
d\theta$\cite{Buttiker}. Therefore, in Fig. 7, we show the results of the spin- and
valley-dependent conductance $G_{\uparrow(\downarrow)}$ and
$G_{K(K^{'})}$ versus $\xi$ for different gate voltages, where
$G_{\uparrow(\downarrow)}=(G_{K\uparrow(\downarrow)}+G_{K^{'}\uparrow(\downarrow)})/2$,
$G_{K(K^{'})}=(G_{K(K^{'})\uparrow}+G_{K(K^{'})\downarrow})/2$. It
is found that some interesting phenomena also exist in the
observable conductance. First, on the whole, with $U$ increasing, the
polarizing effect of the conductance for spin-related electron
within the range of the considered velocity
 barrier becomes more and more obvious, while the polarizing effect of the conductance for valley-dependent electron is becoming weaker as
 shown in Fig.7. Furthermore, concretely we can see that when the gate voltage $U=-2.0\Delta$, the conductance of spin-dependent electron is sensitive to the modulation of Fermi velocity in Fig. 7(a), and there are five points where the situation of polarization inversion occurs. However, the conductance for
 valley-dependent electron is separated completely in the whole range of $\xi$ seen in the Fig. 7(a$^{'}$), especially for velocity barrier
 ratio $\xi=0.6$. Moreover, we can obtain a positive polarization due to the value of conductance for $K$ valley electron being always higher than that
 for $K^{'}$ valley electron.

For $U=-0.2\Delta$, there is only one point of polarization inversion for the spin-dependent electron
in Fig. 7(b), and the corresponding velocity barrier ratio is 1.2. For $\xi<1.2$, a negative polarization can be achieved due to
$G_{\downarrow}>G_{\uparrow}$, while a positive polarization can be obtained for $\xi>1.2$ because the conductance for
spin-down electron sharply declines. In addition, we can find that the polarization for the valley-dependent electron
is not good significantly seen in Fig. 7(b$^{'}$). Particularly, when velocity barrier ratio $\xi>1.0$, the curves of conductance
for two valleys electron are almost overlapped.

When we take the positive voltage $U=0.2\Delta$, the conductance of spin-down electron has nothing to do with the velocity barrier
in Fig. 7(c), and is suppressed completely in the considered range of $\xi$. Therefore, a perfect filtering effect for spin-up
electron can be received. Nevertheless, it is found that the curves of $G_{K(K^{'})}$ are almost the same in the whole range of
velocity barrier ratio seen from Fig. 7(c$^{'}$). Thus, these results provide a practical way for us to manipulate spin and valley currents with different Fermi velocity ratios and voltages in the experiment.

The physical causes of these phenomena may be as follows. Based on the eigenvalue of equation (2)  $E=U+(\tau_{z}s_{z}\lambda-s_{z}h)\pm\sqrt{(\Delta-\tau_{z}s_{z}\lambda)^{2}+(\hbar v_{F}k_{2})^{2}}$, the enhancement of the gate voltage $U$ moves the $K^{бо}$ and $K$ bands upward simultaneously, so that the Fermi energy is shifted to only cross the spin-up bands of $K^{бо}$ and $K$ valley, this results in that for $U=0.2\Delta$ the conductance of spin-down electron is zero, only spin-up electron transports, the perfectly full spin-up polarization can be obtained. Meanwhile it leads to a decline of the valley polarization. The enlargement of velocity barrier makes the Fermi energy move relatively to the band edge of spin-down electron at the $K^{бо}$ and $K$ valley. Therefore, applying the modulation of Fermi velocity, for $U=-2.0\Delta$, the direction of the spin polarization changes complicatedly, while the valley polarization is just positive. For $U=-0.2\Delta$, as the velocity barrier ratio increases, the Fermi energy at the $K^{бо}$ valley only crosses the band of spin-up electron. Thus, the spin polarization has been inverted.

\section{Conclusion}
In this study, by applying velocity modulation, we investigated the
spin- and valley-dependent electron transport properties in a
normal/ferromagnetic/normal monolayer MoS$_{2}$ quantum structure.
To obtain the high spin and valley polarization more efficiently, an
appropriate velocity barrier ratio $\xi$ should be considered in the
structure. The polarization of spin- and valley-dependent electron
can achieve reversal from $100\%$ to $-100\%$ with a gradual
increase in $\xi$. By analyzing the transmission of spin and valley
electron with an increase in the incident angle, we determine that
when the incident angle is larger than $0.16\pi$, the spin and
valley polarizations tends to increase gradually, and the perfect
polarization effect can be achieved at the larger incident angle. In
addition, it is shown that the length $k_{F}L$ and the gate voltage
$U(x)$ of the intermediate ferromagnetic region are also important
in the spin- and valley-dependent electron transport. These
interesting phenomena provide a certain reference value for
experimentally obtaining better electron transport effects using
this model.

This work is supported by NSFC under grants No.11804236,
the General Program of Science and Technology Development Project of
Beijing Municipal Education Commission of China under grants
No.KM201810028005, and Open Research Fund Program of the State Key
Laboratory of Low Dimensional Quantum Physics under grants
No.KF201806.

\end{document}